\input harvmac
\input epsf
\def\lsim{\mathrel{\rlap{\lower4pt\hbox{\hskip1pt$\sim$}}
    \raise1pt\hbox{$<$}}}         
\def\MS{\hbox{$\overline{\rm MS}$}}
\noblackbox
\pageno=0\nopagenumbers\tolerance=10000\hfuzz=5pt
\baselineskip=12pt plus1pt minus1pt
\newskip\footskip\footskip=10pt 
\line{\hfill\tt hep-ph/9806345}
\line{\hfill CERN-TH/98-156}
\line{\hfill DFTT 28/98}
\line{\hfill GeF-TH-6/98}
\vskip 10pt
\centerline{\bf ON POSITIVITY OF PARTON DISTRIBUTIONS}
\vskip 24pt
\centerline{Guido Altarelli}
\vskip 5pt
\centerline{\it Theory Division, CERN, CH-1211 Geneva 23, Switzerland;}
\centerline{\it Universit\`a di Roma Tre, Rome, Italy}
\vskip 10pt
\centerline{Stefano Forte}
\vskip 5pt
\centerline{\it INFN, Sezione di Torino, via P. Giuria 1, I-10125
Torino, Italy}
\vskip 10pt
\centerline{Giovanni Ridolfi}
\vskip 5pt
\centerline{\it INFN,
Sezione di Genova, via Dodecaneso 33, I-16146 Genova, Italy}
\vskip 36pt
{\centerline{\bf Abstract}
\medskip\narrower
\baselineskip=10pt

We discuss the bounds on polarized parton distributions which follow
from their definition in terms of cross section asymmetries. We spell
out how the 
 bounds obtained in the naive parton model can be derived
within perturbative QCD at leading order when all quark and
gluon distributions are defined in terms of suitable physical processes.
We specify a convenient physical definition for the 
polarized and unpolarized gluon
distributions in terms of Higgs production from gluon fusion.
We show that these bounds are modified by subleading
corrections, and we determine them
up to NLO.
We examine the ensuing phenomenological
implications, in particular in view of the
determination of the polarized gluon distribution.\smallskip}
\vskip 48pt

\centerline{Submitted to: {\it Nuclear Physics B}}
\vskip 64pt
\line{CERN-TH/98-156\hfill June 1998}

\vfill\eject \footline={\hss\tenrm\folio\hss}


\def\Tr{\,{\hbox{Tr}}\,}

\def\thru#1{\mathrel{\mathop{#1\!\!\!/}}}

\def\gsim{\mathrel{\rlap{\lower4pt\hbox{\hskip1pt$\sim$}}
    \raise1pt\hbox{$>$}}}         

\def\frac#1#2{{{#1}\over {#2}}}

\def\1{\;1\!\!\!\! 1\;}

\def\smallfrac#1#2{\hbox{${{#1}\over {#2}}$}}

\def\Tr{{\rm Tr}}

\catcode`@=11 
\def\slash#1{\mathord{\mathpalette\c@ncel#1}}
 \def\c@ncel#1#2{\ooalign{$\hfil#1\mkern1mu/\hfil$\crcr$#1#2$}}
\def\lsim{\mathrel{\mathpalette\@versim<}}
\def\gsim{\mathrel{\mathpalette\@versim>}}
 \def\@versim#1#2{\lower0.2ex\vbox{\baselineskip\z@skip\lineskip\z@skip
       \lineskiplimit\z@\ialign{$\m@th#1\hfil##$\crcr#2\crcr\sim\crcr}}}
\catcode`@=12 

\def\PR{{\it Phys.~Rev.~}}

\def\NP{{\it Nucl.~Phys.~}}

\def\PL{{\it Phys.~Lett.~}}
\def\PRep{{\it Phys.~Rep.~}}

\def\ZP{{\it Zeit.~Phys.~}}

\def\vol#1{{\bf #1}}\def\vyp#1#2#3{\vol{#1} (#2) #3}
\newsec{Introduction}

In the naive parton
model, parton distributions are interpreted 
as probability  densities for partons to be
found inside a given target~\ref\guidorep{See e.g. G.~Altarelli,
\PRep\vyp{81}{1982}{1}.}. The partonic viewpoint immediately implies
that if the individual parton distributions are {\it bona fide}
probability densities, they should all be positive. Consequently,
polarized distributions, which are differences of distributions of
partons with polarization parallel or antiparallel to their parent hadron,
should be bounded by the corresponding unpolarized distributions,
which are sums over polarizations. 
Because unpolarized distributions are known rather accurately, such
bounds are potentially useful in the determination of the polarized
distributions, and in particular of the polarized gluon distribution,
which at present can only be indirectly extracted from scaling
violations~\ref\ABFR{G.~Altarelli, R.~D.~Ball, S.~Forte and
G.~Ridolfi, \NP\vyp{B496}{1997}{337}\semi {\tt hep-ph/9803237}.}.
 It is well--known that, 
in perturbative QCD, the naive parton model is
only recovered in the asymptotic limit, when
all perturbative
corrections can be neglected, while at finite energy
the naive parton model results, in general, receive calculable perturbative
corrections. It is thus interesting to ask to which extent the
naive positivity bounds for polarized densities
are modified by QCD corrections,
especially in the case of the gluon distribution, which is of special
phenomenological interest.

Here we  address this question
by deriving bounds on the polarized quark and gluon distributions
from positivity of cross sections. We show that  the naive positivity bounds
can be derived from leading order (LO) QCD, provided that the quark
and gluon distributions are defined in terms of polarized
and unpolarized physical processes. 
While it is clear that an adequate choice of defining process for
quark densities is in terms of deep--inelastic structure functions,
such as, e.g., $F_2$ and $g_1$, the selection of a suitable process for the
gluon is subtler. In fact, in order to directly obtain the naive bound
on $|\Delta g|/g$ at LO, we need a process where only gluons of a
given polarization contribute at LO. Thus, one is led to consider the
production of a scalar particle (such as the Higgs)
in gluon--proton collisions: only the
gluon in the proton contributes at LO, and, moreover, if the external
gluon and the target are polarized, only one polarization state for the
gluon parton is selected.

Next, we discuss how these bounds get modified by 
next-to-leading order (NLO) corrections, in general in a
scheme-dependent way. In
a generic scheme (such as 
the commonly used \MS\
scheme) the NLO bounds may be more or less restrictive than the LO
ones, depending on the parton distribution and the region of $x$ which
are considered. We  then
determine the full set of NLO bounds for quark and gluon partons
explicitly. This requires the determination of the NLO corrections
in the polarized case
to the Higgs production process which defines the gluon at LO.
After having established the improved NLO bounds, we use them to extract
phenomenological information on polarized parton distributions in
terms of the unpolarized ones, and
specifically to constrain the polarized gluon distribution.

\newsec{Positivity of parton distributions at leading order}

In perturbative QCD, positivity bounds on parton distributions follow
from their definitions in terms of cross-section asymmetries, and 
from
positivity of physical cross sections. 
Let us first consider the simple case
of the structure function $g_1(x,Q^2)$
for inclusive deep-inelastic scattering
of longitudinally polarized leptons on a longitudinally polarized
target,  which leads to bounds on polarized quark and
antiquark distributions. Neglecting power corrections, $g_1$
is related to the
asymmetry $A_1$ for deep-inelastic
scattering of transversely
polarized virtual photons on a longitudinally polarized nucleon 
through~\ref\mauro{See e.g. M.~Anselmino, A.~Efremov and E.~Leader,
\PRep\vyp{261} {1995}{1}.}
\eqn\aone{A_1\equiv
{\sigma_{1/2}-\sigma_{3/2}\over\sigma_{1/2}+\sigma_{3/2}}={g_1(x,Q^2)\over
F_1(x,Q^2)},}
where the subscripts denote the total angular momentum of the
photon-nucleon pair along the incoming lepton's direction.
This immediately implies that
$g_1$ is bounded by its unpolarized
counterpart $F_1$:
\eqn\boundx{|g_1(x,Q^2)|\leq
F_1(x,Q^2).}

The structure functions $F_1$ and $g_1$ are related to parton
 distributions through coefficient functions according to
\eqnn\fonedef\eqnn\gonedef
$$\eqalignno{
F_1(x,Q^2)&={{1}\over{2}}
\sum_{i=1}^{n_f} e^2_i C^d_i\otimes \left(q_i+\bar q_i\right)+ 
2n_f \langle e^2\rangle C^d_g \otimes 
g&\fonedef
\cr
g_1(x,Q^2)&={{1}\over{2}}
\sum_{i=1}^{n_f} e^2_i \Delta C^d_i\otimes \left(\Delta q_i+\Delta
\bar q_i\right)+
2n_f\langle e^2\rangle\Delta C^d_g \otimes \Delta
g,&\gonedef\cr}$$
where
$\otimes$ denotes the usual convolution with respect to
$x$, $q_i$ and $\bar q_i$ are  quark
and antiquark distributions of flavour $i$ and with electric charge
$e_i$,  $\langle e^2\rangle\equiv {1\over n_f}\sum_{i=1}^{n_f} e^2_i$,
$g$ is the
 gluon distribution, 
$\Delta q_i$, $\Delta \bar q_i$ and $\Delta
g$ the corresponding polarized distributions, and 
$\Delta C^d$ and
$C^d$ are, respectively, polarized and unpolarized coefficient
functions
(the index $d$ is a reminder that these are deep-inelastic
coefficient functions). 
The coefficient functions
can be written as series in $\alpha_s$ with $x$-dependent
coefficients: 
\eqn\cexp{C(x,\alpha_s)=\sum_{k=0}^\infty 
\left(\alpha_s \over 2\pi\right)^k
C^{(k)}(x).}

Now, at LO 
\eqn\locfq{\eqalign{
&C^{d,\,(0)}_{i}(x)=\Delta
C^{d,\,(0)}_{i}(x)=\delta(1-x)\cr
&C^{d,\,(0)}_{g}(x)=\Delta C^{d,\,(0)}_{g}(x)=0,\cr}} 
so that at LO eq.~\boundx\ immediately implies
\eqn\boundpdf{
\left|
\sum_{i=1}^{n_f}e^2_i\left[ \Delta q_i(x,Q^2)+\Delta \bar
q_i(x,Q^2)\right]\right |
\leq \sum_{i=1}^{n_f}e^2_i
\left[ q_i(x,Q^2)+ \bar q_i(x,Q^2)\right].}
The bound eq.~\boundpdf\ must be
satisfied as a matter of principle
for any choice of target, i.e. for any
combination of quark plus antiquark distributions, and thus it must be
satisfied by each quark flavour separately. Furthermore, the bound
must also be satisfied by the structure functions for charged--current
scattering, and thus by each quark and antiquark distribution separately.
Therefore, we conclude that at  LO 
\eqn\ibound{|\Delta q_i(x,Q^2)|\leq
q_i(x,Q^2)} 
for all flavours $i$, for all $x$, and for all $Q^2$ such that the LO
approximation makes sense.

The constraint eq.~\boundpdf\ immediately implies that 
at LO it is possible to interpret the polarized and
unpolarized quark distributions according to a naive parton picture.
Namely, if we define distributions of
partons polarized parallel or antiparallel to their parent hadron by
\eqn\fnfq{
q^\pm_i={1\over2}(q_i\pm\Delta q_i),\quad
\bar q^\pm_i={1\over2}(\bar q_i\pm\Delta \bar q_i),
}
then eq.~\ibound\ implies that $q^\pm_i$  and $\bar q^\pm_i$ are
positive--(semi)definite  and can thus be
interpreted as probability distributions. 

Next, one may ask whether a LO positivity bound can likewise be
derived for the gluon distribution.  In order to do so, we must select
a pair of polarized and unpolarized processes which at LO define the
gluon, i.e.  processes whose LO coefficient functions are in an
analogous way
given by
\eqn\locfg{\eqalign{
&C^{h,\, (0)}_{i}(x)=\Delta
C^{h,\, (0)}_{i}(x)=0\cr
&C^{h,\, (0)}_{g}(x)=\Delta C^{h,\, (0)}_{g}(x)=\delta(1-x).\cr}} 
Because the
gluon only couples strongly, the leading nontrivial order must be at least
$O(\alpha_s)$, unlike the quark which can be defined at
$O(\alpha_s^0)$.

Processes which are typically used to determine the
gluon distribution experimentally by singling out the 
photon--gluon fusion diagram, such as
heavy quark production, are not useful to our present 
purpose, because even
though they do have the property that at LO only the gluon coefficient
functions are nonvanishing, they do not satisfy the requirement that
the spin--flip coefficient function
vanishes, i.e. $C_g=\Delta C_g$. Therefore, with this definition
positivity of the physical cross sections does
not automatically imply positivity of the parton distribution. 
One could also think of using
deep-inelastic scattering mediated by longitudinally polarized
photons, i.e. in the unpolarized case the structure function $F_L$,
which receives a LO gluon
contribution (of order $\alpha_s$)~\ref\fl{A.~Zee, F.~Wilczek and S.~B.~Treiman, {\it
Phys. Rev.} 
{\bf D10} (1974) 2881.} (of order $\alpha_s$. However, 
there is no suitable  polarized
counterpart of $F_L$.

\topinsert
\vbox{
\hfil\epsfysize=6cm\epsfbox{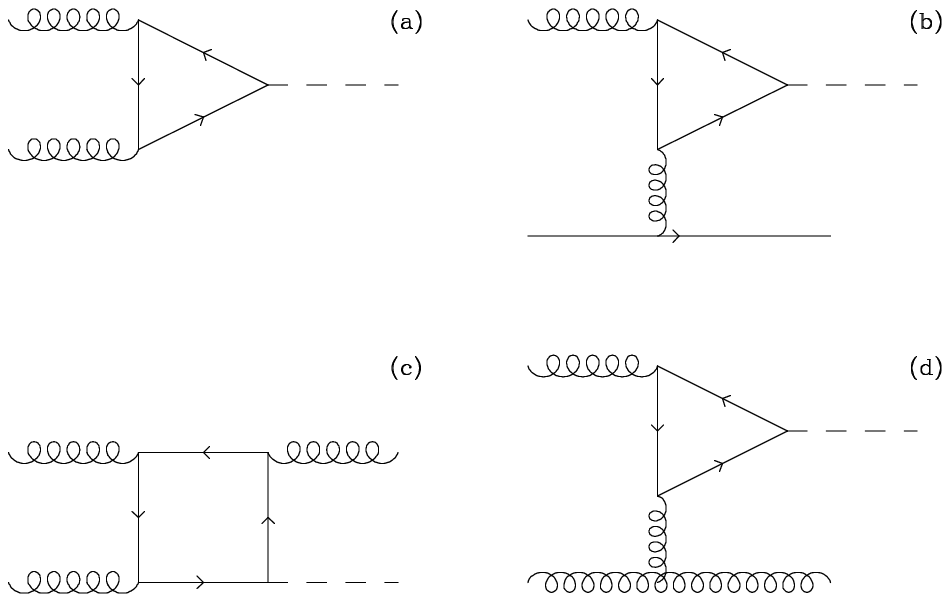}\hfil}
\bigskip\noindent{\footnotefont\baselineskip6pt\narrower
Figure 1: Diagrams for the processes (a)  $g+g\to H$,
(b) $g+q\to H+q$, (c, d) $g+g\to H+g$.
\medskip}
\endinsert
Nevertheless, the bound
\eqn\gbound{|\Delta g(x, Q^2)|\leq g(x,Q^2).}
is directly obtained at LO if we  
define the gluon distribution from inclusive  Higgs
production in gluon--proton scattering,
i.e. from the process $g+p\to H+X$. At leading nontrivial order
[i.e. $O(\alpha^2_s)$] this
process proceeds through $g+g\to H$, induced by  coupling of the
gluons to a top quark loop (see fig.~1a). Even though this process is
not realistic for an experimental measure of 
the gluon distribution,
it is particularly convenient to formally define the gluon
distribution and establish positivity bounds.
The cross section for the LO partonic unpolarized process is given
by~\ref\loh{F.~Wilczek, {\it Phys. Rev. Lett.} {\bf 39} (1977) 1304.}
\eqn\lohxs{\sigma(gg\to H)= {\alpha_s^2(m_h^2)G_Fm_h^2\over 
288\sqrt{2}
\pi}|A|^2\delta(s-m_h^2),}
where $s$ is the center-of-mass energy of the $gg$ collision, and $A$
is a dimensionless
function of the Higgs mass and the top
quark mass: $A=A(m_t^2/m_h^2)$,
such that $\lim_{r\to\infty}|A(r)|^2=1$.
Furthermore,
since the Higgs particle is scalar, only the polarized
amplitude in which the two gluons have the same helicity is nonzero at
LO in any frame where the two gluon momenta are in opposite
directions. 

We can thus define the polarized and unpolarized gluon
distributions by considering the gluon--proton scattering process.
It is important to note that the external gluon is to be treated
as a hadron, in the sense that it must be endowed with its own
nonperturbative parton densities $g_g$, $\Delta g_g$, $u_g$, $\Delta
u_g$ and so forth. The collinear singularities associated with the
initial gluon leg have then to be absorbed into a redefinition of the
associated parton distributions. We are however free to choose the
resulting distributions in a form which is convenient for our
conceptual purposes. Our choice corresponds to that of an effective
free gluon, i.e. $g_g(x)=\Delta g_g(x)=\delta(1-x)$, $u_g=\Delta
u_g=\dots =0$.
As a consequence, the gluon--proton cross section
can be written as
\eqn\lohfc{\eqalign{
{1\over x}\sigma[gp\to H+X](x,m_h^2)&={\alpha^2_s(m_h^2)G_F\over 288\sqrt{2}
\pi}|A|^2\int_x^1\!{dy\over y} \left\{C^h_g\left({x\over y}\right)
g(y,m_h^2)\right.\cr&\qquad\qquad\left.
+C^h_s\left({x\over y}\right)\Sigma(y,m_h^2)\right\}\cr
{1\over x}\Delta\sigma[gp\to H+X](x,m_h^2)&={\alpha^2_s(m_h^2)G_F\over 288\sqrt{2}
\pi}|A|^2\int_x^1\!{dy\over y} \left\{\Delta C^h_g\left({x\over y}\right)
\Delta g(y,m_h^2)\right.\cr&\qquad\qquad\left.+\Delta C^h_s\left({x\over y}\right)
\Delta \Sigma(y,m_h^2)\right\},\cr}}
where $x={m_h^2\over s}$ and the singlet quark distribution is defined
as
\eqn\sidef{ \Sigma(x,Q^2) =\sum_{i=1}^{n_f} \left[q_i(x,Q^2)+\bar
 q_i(x,Q^2)\right],}
and likewise for the polarized distribution $\Delta \Sigma$.

The unpolarized and polarized cross sections
$\sigma$ and $\Delta \sigma$
are respectively the sum and the difference of the cross sections 
with fixed  parallel and antiparallel proton and gluon helicity.
We have included a contribution from the scattering of the
incoming gluon off quark and antiquark partons in the proton, which is
present at higher orders even though it vanishes at LO where only
the diagram of fig.~1a contributes.
If we then expand  the coefficient functions
(i.e. rescaled partonic cross sections) $C^h$ and $\Delta C^h$ 
in powers of $\alpha_s$ it follows immediately that the LO term in the
expansion is given by eq.~\locfg. Positivity of the cross sections with
fixed gluon and proton helicities then implies 
\eqn\gboundx{|\Delta
\sigma(x,m_h^2)|\leq\sigma(x,m_h^2)} 
and thus the LO bound eq.~\gbound.
We conclude that the naive partonic 
positivity bound eqs.~\ibound,\gbound\ hold for quarks and gluons 
provided the scale is large
enough that NL and higher order corrections are negligible.

Since parton distributions themselves 
are scale dependent, one is led to ask
how the positivity bounds will vary with scale at
LO~\ref\lead{C.~Bourrely, E.~Leader and O.~V.~Teryaev, {\tt
hep-ph/9803238}.}. 
The scale dependence of parton distributions is expressed in terms of
the singlet distribution eq.~\sidef\ and 
nonsinglet quark distributions defined as
\eqn\nsdef{ q_{\rm NS}(x,Q^2)=\sum_{i=1}^{n_f}
\left(\smallfrac{e_i^2}{\langle e^2\rangle}-1\right)
( q_i(x,Q^2)+\bar q_i(x,Q^2)),}
and likewise for the polarized distributions $\Delta q_{\rm NS}$.
The moments of the
nonsinglet distributions then evolve multiplicatively:
\eqn\nsev{q_{\rm NS}(N,Q^2)=
\left({\alpha_s(Q_0^2)\over\alpha_s(Q^2)}\right)^{a_{\rm NS}(N)}q_{\rm
NS}(N,Q_0^2),}
where for a generic distribution $f(x)$ we have defined
$f(N)=\int_0^1 dx x^{N-1} f(x)$.
Moments of the singlet 
quark evolve by mixing with the gluon, according to
\eqn\sev{\eqalign{&\left[c^1_\pm(N) \Sigma (N,Q^2)+c^2_\pm(N)g(N,Q^2)\right]=
\cr&\qquad\qquad=
\left({\alpha_s(Q_0^2)\over\alpha_s(Q^2)}\right)^{a_{\pm}(N)}
\left[c^1_\pm(N) \Sigma (N,Q_0^2)+
c^2_\pm(N)g(N,Q_0^2)\right],\cr}} 
and likewise for polarized distributions. 

Now, the LO
anomalous dimensions (which are the only relevant ones asymptotically)
satisfy  $a_+(N)\geq a_{NS}(N)\geq a_-(N) $  as it is
trivial to verify from their explicit expression~\guidorep. It follows
that the largest singlet anomalous dimension $a_+(N)$ will
asymptotically drive the
evolution of all the quark distributions for each flavour and of the
gluon distribution. The same of course applies to  polarized anomalous
dimensions $\Delta a(N)$. It is also  easy to
see that necessarily $a_+(N)\geq| \Delta a^+(N)|$ for all $N$: this is an
immediate consequence of the fact that LO unpolarized and polarized
anomalous dimensions are respectively sums and differences of
the square moduli of invariant particle emission vertices~\ref\ap{
G.~Altarelli and G.~Parisi, \NP\vyp{B126}{1977}{298}.} (up to
virtual contributions which are the same in the polarized and
unpolarized case).  Indeed, it is easy to check 
from their explicit
expressions~\guidorep\ that
for all finite $N$, $a_+(N)>|\Delta a_+(N)| $, while
$\lim_{N\to\infty}  a_+(N)= \lim_{N\to\infty} \Delta a_+(N)$).
 It follows
that for all finite $N$ 
\eqn\limev{\lim_{Q^2\to\infty}{|\Delta q_i(N,Q^2)|\over q_i(N,Q^2)}=
{|\Delta g(N,Q^2)|\over g(N,Q^2)}=0.
}

Hence, the LO bounds eqs.~\ibound,\gbound\ will always be
satisfied at large enough scale. Notice that, conversely, this implies
that the LO bounds will always be violated at sufficiently low scale,
or, otherwise stated, the (scale--independent) LO bounds become more
and more restrictive as the scale is decreased, i.e. they become
unacceptably restrictive if imposed at an excessively low scale.
On the other hand,
this raises the issue of computing subleading corrections to the bounds.
 We will do this in the next section.

\newsec{Positivity beyond leading order}

At NLO and beyond helicity--flip cross
sections are no longer zero, so that
the polarized and unpolarized coefficient
functions $C$ and $\Delta C$
are no longer equal to each other. In addition, the gluon
DIS coefficient function $C_g^d$ and the quark Higgs production
coefficient function $C_s^h$ become nonzero. 
As a consequence, the LO positivity bounds, eqs.~\ibound,\gbound\
receive NL corrections. 
In order to discuss these corrections, it is convenient to first
rewrite the structure functions eqs.~\fonedef-\gonedef\ by separating
the quark singlet and nonsinglet, rather than individual flavours, since the
NLO bounds are affected by quark--gluon mixing, but this only involves the
singlet quark distribution:
\eqnn\fonesns\eqnn\gonesns
$$\eqalignno{
F_1(x,Q^2)&=\smallfrac{\langle e^2 \rangle}{2} [C^d_{NS}\otimes q_{NS}
+C^d_s\otimes  \Sigma + 2n_fC^d_g \otimes  g],
&\fonesns
\cr
g_1(x,Q^2)&=\smallfrac{\langle e^2 \rangle}{2} [\Delta 
C^d_{NS}\otimes \Delta q_{NS}
+\Delta C^d_s\otimes \Delta \Sigma + 2n_f\Delta C^d_g \otimes \Delta g].
&\gonesns\cr}$$

The LO positivity bounds on individual quark and antiquark flavours
eq.~\ibound\  then imply LO positivity of the C-even combination
\eqn\cpbound{|\Delta q_i(x,Q^2)+\Delta \bar q_i(x,Q^2)|\leq
\left[q_i(x,Q^2)+\bar q_i(x,Q^2)\right],} 
which in turn implies LO positivity of the singlet quark distribution,
eq.~\sidef\
\eqn\sibound{|\Delta \Sigma(x,Q^2)|\leq
\Sigma (x,Q^2).}
This is of course a necessary, but not sufficient condition for
LO positivity of individual flavours, eq.~\cpbound, to hold. A necessary
and sufficient condition for eq.~\cpbound\ is given by constraining
the polarized singlet by
eq.~\sibound, and the polarized nonsinglet through constraints
which depend both on the value of $\Delta \Sigma$ and on the
unpolarized (singlet and nonsinglet) distributions. The construction
of such constraints is trivial: consider for instance the simple
two--flavour case. The constraints eq.~\cpbound\ then mean that a
rectangular area is allowed in the ($\Delta u$,~$\Delta d$) plane. The
corresponding equivalent constraints on $\Delta \Sigma$ and $\Delta
q_{NS}$ are found by giving coordinates in the plane with respect
to the $\Delta \Sigma$--$\Delta
q_{NS}$ axes, which are rotated by $\pi/4$ with respect to the original
axes.  Likewise, for the separate positivity of quarks and antiquarks
(eq.~\ibound):
necessary and sufficient conditions can be derived by imposing positivity of the
C--even combination, eq.~\cpbound, and a similarly derived condition
on the C--odd combination $\Delta q_i-\Delta \bar q_i$.

We are now ready to discuss the NLO modification of the
quark and gluon bounds eqs.~\ibound,\gbound. First, note that the bounds on
physical cross sections eqs.~\boundx-\gboundx\ 
must be satisfied for all $x$, and therefore also by the corresponding
Mellin moments:
\eqnn\boundn\eqnn\gboundn
$$\eqalignno{|g_1(N,Q^2)|&\leq
F_1(N,Q^2)&\boundn\cr
|\Delta \sigma (N,Q^2)|&\leq
\sigma (N,Q^2).&\gboundn\cr}$$
Since the bound eq.~\boundn\ must hold for any target, the same
arguments which lead from eq.~\boundpdf\ to eq.~\sibound\ now give the
general bound on the singlet quark distribution
\eqn\genqbound{\eqalign{
\left|\Delta C^d_s(N,Q^2)  \Delta \Sigma(N,Q^2) + 2n_f\Delta C^d_g(N,Q^2)
\Delta g(N,Q^2)\right| &\cr
\leq\big[ C^d_s(N,Q^2)  \Sigma(N,Q^2) + 2n_f&C^d_g(N,Q^2)   g(N,Q^2)\big],
\cr}}
which at NLO implies
\eqn
\nloqbound{
{\left|\Delta\Sigma(N,Q^2)
      \left(1+{\alpha_s\over 2\pi}\Delta C^{d,\,(1)}_s(N,Q^2)\right)
      +2n_f{\alpha_s\over 2\pi} 
\Delta C^{d,\,(1)}_g(N,Q^2)\Delta g(N,Q^2)\right| \over
\Sigma(N,Q^2)
      \left(1+{\alpha_s\over 2\pi} C^{d,\,(1)}_s(N,Q^2)\right)
      +2n_f{\alpha_s\over 2\pi} 
 C^{d,\,(1)}_g(N,Q^2) g(N,Q^2)
}\leq 1}
Similarly, eq.~\gboundn\ gives the NLO bound on the gluon
distribution
\eqn\nlogbound{{
\left| \Delta g(N,Q^2)\left(1+ {\alpha_s\over 2\pi} \Delta 
C^{h,\,(1)}_g(N,Q^2)\right)+{\alpha_s\over 2\pi} \Delta 
C^{h,\,(1)}_s(N,Q^2)
\Delta \Sigma(N,Q^2)\right| \over
  g(N,Q^2)\left(1+ {\alpha_s\over 2\pi}  
C^{h,\,(1)}_g(N,Q^2)\right)+{\alpha_s\over 2\pi} 
C^{h,\,(1)}_s(N,Q^2)
\Sigma(N,Q^2)
}\leq 1}

Equivalently, we may reabsorb the NL modification of the bounds in a
change of factorization scheme. Indeed, we are always free to choose 
to work in a ``parton'' scheme in which the coefficient functions for
the defining process,
computed to any given perturbative order,
coincide with their LO expressions
\eqn\lonqg{ \eqalign{
&C^{d}_s(N)=C^{h}_g(N)=\Delta C^{d}_s(N)=
\Delta C^{h}_g(N)=1\cr
&C^{d}_g(N)=C^{h}_s(N)=\Delta C^{d}_g(N)=
\Delta C^{h}_s(N)=0.\cr}}
Imposing that all higher order
terms vanish
$C^{d,\,(i)}=C^{h,\,(i)}=\Delta C^{d,\,(i)}=\Delta C^{h,\,(i)}=0$,
$i\ge1$ determines the factorization scheme completely: 
given coefficient functions up to, say, NLO in any  scheme (such as 
the \MS\ scheme) we can transform to the parton scheme by redefining the
singlet quark and gluon distributions according to
\eqn\schch{
\pmatrix{\Sigma^{\rm parton}(N,Q^2) \cr g^{\rm parton}(N,Q^2)}
=\left[\1+\frac{\alpha_s(Q^2)}{2\pi}
\pmatrix{C^{d,\,(1)}_s(N)& C^{d,\,(1)}_g(N) \cr 
C^{h,\,(1)}_s(N)& C^{h,\,(1)}_g(N) }
\right] \pmatrix{ \Sigma(N,Q^2) \cr  g(N,Q^2)},}
and likewise for polarized parton 
distributions.
Otherwise stated, we are always free to choose a scheme in which the
bounds reduce to their naive LO form. However, in
a generic scheme parton distributions at NLO and beyond
will not necessarily respect the LO positivity constraints 
eqs.~\ibound,\gbound.
Nevertheless, once we impose the correct positivity  bounds
eq.~\nloqbound-\nlogbound\ on  NLO parton distributions in any scheme
of our choice,
physical cross sections will be positive up to NNLO corrections, and
if the parton distributions are transformed to any other scheme they
will respect the positivity bound pertinent to that scheme.

The coefficient functions $C^d$, $\Delta C^d$ and $C^h$ are all known
up to NLO.
In order to determine the full set of NLO positivity bounds
and study their phenomenological implications we must thus still
determine the NLO polarized quark and gluon coefficient functions $\Delta C^h$.
We will do this in the next section.

\newsec{Determination of the polarized
Higgs production coefficient functions to next-to-leading order}

At NLO the process $g+p\to H+X$  can also proceed through
gluon--quark fusion, $g+q\to H+q$ (fig.~1b), so the quark coefficient
functions $C_q^h$, $\Delta C_q^h$ are nonzero; furthermore since there
is an extra particle in the final state both helicity configurations
can now contribute and the polarized and unpolarized coefficient
functions are unequal. Similarly, the gluon--gluon fusion
contribution can now proceed with emission of an extra gluon, $g+g\to
H+g$ (figs.~1c,d), so that the gluon polarized and unpolarized coefficient
functions are likewise unequal. The LO process finally receives
virtual corrections which necessarily have  the same momentum
and spin dependence of the LO, and cancel IR divergences of the real
emission contributions. 
All these contributions take a simple form when calculated in
the limit ${m_t\over m_h}\to
\infty$, in which the quark loop behaves as
a pointlike effective interaction~\ref\vain{A.~Vainshtein, V.~Zakharov
and M.~Shifman, {\it Sov. Phys. Usp.} {\bf 23} (1980) 429.}. We will
determine the polarized cross section at NLO within this
approximation, which, regardless of its phenomenological relevance, is
adequate for conceptual purposes. The corresponding
unpolarized calculations have been performed in ref.~\ref\kis{R.~K.~Ellis
et al., {\it Nucl. Phys.} {\bf B297} (1988) 221.} for the real
emission diagrams while the virtual corrections have been determined  
in ref.~\ref\rvirt{S.~Dawson, {\it Nucl. Phys} {\bf
B359} (1991) 283\semi A.~Djouadi, M.~Spira and P.~M.~Zerwas, {\it
Phys. Lett.} {\bf B264} (1991) 440.}.

The polarized parton cross sections which define the coefficient
functions eq.~\lohfc\ are determined by computing helicity differences
of squared matrix elements, i.e.,  for the emission process $g+f\to H+f$,
with $f=q,\,g$ we have
\eqn\amp{\Delta M_f(s,t,u)\equiv |M_f^{\uparrow\uparrow}(s,t,u)|^2-
|M_f^{\uparrow\downarrow}(s,t,u)|^2,}
where $s$, $t$ and $u$ are the Mandelstam invariants of the process
and the arrows refer to the helicities of the incoming $g$ and $f$ partons.
This difference can in turn be determined by
contracting the tensor amplitudes with polarized projectors (more
precisely, differences of projectors):
for the real gluon emission process $g+g\to H+g$
\eqn\gamp{\Delta M_g(s,t,u)=
-M_g^{\mu\nu\rho}M_g^{*\,\alpha\beta}{}_\rho P^g_{\mu\alpha} P^g_{\nu\beta},}
where the gluon projector is given by
\eqn\gluproj{P^g_{\mu\nu}=i\epsilon_{\mu\nu\rho\sigma} {n^\rho
k^\sigma\over n\cdot k}}
in terms of the momentum $k$ carried by the gluon line
on which the projector is
applied and an
arbitrary 
lightlike 
vector $n$ such that $k\cdot n \not = 0$
(for example, the momentum of the other gluon
line).
For the quark emission process $g+q\to H+q$ we have 
\eqn\qamp{\Delta M_q(s,t,u)=
\Tr \left[M_q^\mu P^q\thru k M_g^\dagger{}^\nu\thru k^\prime 
\right]P_{\mu\nu}^g,}
where the trace runs over fermion indices, $k$ and $k^\prime$ are the
incoming and outgoing quark momenta,  and the quark projector on
the incoming line is given by 
\eqn\quproj{P_q=-\gamma_5.}

We can now compute $\Delta M_f$ in dimensional regularization
in $d$ dimensions; the calculation closely parallels that of the
unpolarized contribution~\refs{\kis-\rvirt}. The polarized calculation
requires a recipe to handle
 the $\gamma_5$ matrix and the antisymmetric tensor: we will 
continue them to $d$ dimensions according to the HVBM~\ref\hvbm{
G.~'t~Hooft and M.~Veltman, {\it Nucl. Phys} {\bf B44} (1972) 189\semi
P.~Breitenlohner and D.~Maison, {\it Comm. Math. Phys.} {\bf 52}
(1977) 11.} prescription, since this will give results in the same
scheme in which the NLO anomalous dimensions of all parton
distributions have been determined~\ref\ucc{
R.~Mertig and W.~L.~van~Neerven, \ZP\vyp{C70}{1996}{637}\semi
              W.~Vogelsang, \PR\vyp{D54}{1996}{2023},
\NP\vyp{B475}{1996}{47}.}. 
In the gluon case the tensor amplitudes are given in ref.~\kis, while
in the quark case we can readily calculate the amplitude $M_q^\rho$
from the diagram of fig.~1b. We get\eqnn\sqampg\eqnn\sqampq
$$\eqalignno{\Delta M_g(s,t,u)&=\alpha_s^3G^d_F {64\over 3\sqrt{2}\pi}  {m_h^8+s^4-t^4-u^4\over s t u}
& \sqampg\cr
\Delta M_q(s,t,u)&=\alpha_s^3G^d_F {64\over 3\sqrt{2}\pi}{s^2-u^2\over 2 t} ,
&\sqampq\cr}$$
where $G^d_F$ is the $d$--dimensional continuation  of the strength of
the effective $ggH$ interaction~\rvirt. The form of this continuation
is immaterial because it appears as an overall factor in all the
separately divergent contributions which add up to the final
finite NLO result.

The kinematics for the inclusive process can be entirely specified in
terms of $m_h^2$, $s={m_h^2\over x}$,  
and the center-of-mass scattering angle of the
emitted parton $\cos \theta= 1-2y$, in terms of which the remaining
Mandelstam invariants are given by
\eqn\kin{\eqalign{t&=-(s-m_h^2) (1-y)=-m_h^2 (1-y){1-x\over x}\cr
u&=-(s-m_h^2) y=-m_h^2y{1-x\over x} .\cr}}
The physical cross-section is obtained by integrating the real
emission amplitude
over phase space, adding vitual corrections and collinear and
ultraviolet counterterms, 
and supplying the appropriate flux factor:
\eqn\genxs{\Delta \sigma(g+f\to H+f)={1\over 2 s}
\int\!d\phi_2 \,\Delta M_f
+\Delta \sigma_f^{\rm
virt}+\Delta \sigma_f^{\rm coll}+\Delta \sigma^{uv}_f.}
The two-body phase space in $d=2(2-\epsilon)$ dimensions
is given by
\eqn\phsp{d\phi_2={1\over 8\pi}{1\over\Gamma(1-\epsilon)}
\left(4\pi\over s\right)^\epsilon
\left(1-{m_h^2\over s}\right)^{1-2\epsilon} y^{-\epsilon}(1-y)^{-\epsilon} dy.}

The collinear counterterms remove the singularities which correspond
to collinear radiation of the final--state parton by the incoming one,
and are thus given by
\eqn\coll{\Delta \sigma^{\rm coll}_f=\int_x^1\!{dy\over y} 
\Delta \sigma_0(m_h^2,y) {1\over \epsilon}
\Gamma(1+\epsilon) (4\pi)^\epsilon P_{gf}\left({x\over y}\right)}
where $P_{gf}$ is the LO QCD splitting function ~\ap, the
factor of $\Gamma(1+\epsilon) (4\pi)^\epsilon$ accompanying the
$\epsilon$ pole characterizes the \MS\ subtraction scheme, and
$\Delta \sigma_0(m_h^2,x=m_h^2/s)$ is the polarized Born
cross section in $d$ dimensions, which turns out to coincide with the
unpolarized four--dimensional cross-section eq.~\lohxs\ up to the
replacement $G_F\to G_F^d$.\foot{Notice that the fact that the
dependence on $\epsilon$ of $\Delta \sigma_0$ only comes through
$G_F^d$ is nontrivial, and peculiar of the polarized calculation where
it follows from the adoption of the HVMB prescription for the
antysimmetric tensor.}
Virtual corrections, as well as the ultraviolet counterterm related to
charge renormalization, only contribute to the gluon process, since the
quark process only starts at NLO, and are proportional to the Born
cross section in $d$ dimensions. They were computed in ref.~\rvirt\
from which they can be simply obtained by replacing the $d$
dimensional unpolarized  Born cross section with its polarized
counterpart, and are thus given by
\eqn\virt{\eqalign{
\Delta \sigma^{\rm virt}_g=\Gamma(1+\epsilon) (4\pi)^\epsilon
\Delta \sigma_0(m_h^2,x)\left({3\over\epsilon^2}-{11\over2}-2\pi^2\right)
&\cr
\Delta \sigma^{\rm uv}_g=
\Delta \sigma_0(m_h^2,x) {11-2/3 n_f\over 4\pi}{1\over 2\epsilon}
\Gamma(1+\epsilon) (4\pi)^\epsilon
&.\cr}}

Performing the phase--space integral eq.~\phsp\ of the square
amplitudes eq.~\sqampq-\sqampg\ and adding the collinear
counterterms~\coll\ and, in the gluon case, the virtual corrections
eq.~\virt\ we  get the physical cross sections for the NLO
processes $g+g\to H+g$ and $g+q\to H+q$. Double poles in $\epsilon$ cancel
between real and virtual contributions to the gluon process. Poles 
in $\epsilon$ proportional to $\delta(1-x)$
are generated in the real contribution to
the gluon process because of the presence of a
term proportional
to $(1-x)^{-(1+2\epsilon)}$ which has an IR singularity in the
$\epsilon\to0$ limit:
\eqn\irreg{{1\over (1-x)^{1+2\epsilon}}=
-{1\over 2\epsilon}\delta(1-x)
+{1\over (1-x)_+}
-2\epsilon\left({\ln(1-x)\over1-x}\right)_+
+{\cal O}(\epsilon^2).}
These poles cancel against those in the virtual correction eq.~\virt.
The collinear simple
poles are cancelled by the splitting function counterterms in both the
quark and gluon emission diagrams. 

The final finite result for the
physical cross sections gives us immediately the coefficient functions
$\Delta C^h$ eq.~\lohfc\ by dividing out the physical
Born cross section eq.~\lohxs:
\eqnn\nlresxg\eqnn\nlresxq
$$\eqalignno{&\Delta C^{h,\,(1)}_g(x)=11{(1-x)^3\over x}
+\left(11+2\pi^2\right)\delta(1-x)\cr&\qquad\qquad+2
\left[{\ln(1-x)^2\over1-x}\right]_+\Delta R_{gg}(x)-
2 \ln x \,\Delta P_{gg}(x)
\cr&\qquad\qquad\qquad\qquad+
2\ln\left({m^2_h\over \mu^2}\right)\Delta P_{gg}(x)
&\nlresxg\cr
&\Delta
C^{h,\,(1)}_s(x)=2{(1-x)^2\over x}+
\Delta P_{gq}(x)\left[\ln{(1-x)^2\over x}
+\ln\left({m^2_h\over \mu^2}\right)\right],
&\nlresxq\cr}$$ 
where $\Delta P_{fg}$ are LO QCD polarized
splitting functions, 
\eqn\rggdef{{\Delta R_{gg}(x)\over (1-x)_+}=
\left[\Delta P_{gg}(x)-{\beta_0\over2} \delta(1-x)\right],} 
$\beta_0=11-{2\over 3} n_f$ is the leading
coefficient of the QCD beta function, and $\mu^2$ is a
factorization scale.
We can now derive the full set of positivity bounds for the quark and
gluon distribution. The result eq.~\nlresxg-\nlresxq\ is of course
also in principle interesting for its own sake as a  determination of
the cross section for 
Higgs production in polarized gluon-proton 
scattering.\foot{Note that eqs.~\nlresxg-\nlresxq\ give the
coefficient of $\alpha_s\over 2\pi$ in the perturbative expansion of
the coefficient function. 
The coefficient functions eqs.~\nlresxg-\nlresxq\  are thus related 
to the $O(\alpha_s^3)$ contributions to the 
cross section for the process $g+f\to H+X$ by
$\Delta \sigma(g+f\to H+X)={\alpha^3_s G_F\over576 \sqrt{2}\pi^2}
xC^{h,\,(1)}_f(x)$.}

\newsec{Phenomenology of next-to-leading order positivity bounds}

We can now collect all results for the eight coefficient functions
which determine the NLO positivity bounds for quarks and gluons,
eqs.~\nloqbound-\nlogbound, in moment space. The $F_1$ coefficient
functions \MS\ can be determined from the known
$F_2$~\ref\ftwo{W.~A.~Bardeen et al, {\it Phys. Rev.} {\bf D18} (1978)
3998.} 
and $F_L$~\fl\ 
coefficient functions using $F_1=(F_2-F_L)/(2x)$. The $g_1$
coefficient functions are  available~\ref\kod{J.~Kodaira, {\it
Nucl. Phys.} {\bf B165} (1980) 129.} in the \MS\ scheme with HVBM
prescription; from these it is straightforward to determine them in
the AB version of the \MS\ scheme~\ref\bfr{R.D.~Ball, S.~Forte and
G.~Ridolfi, \PL\vyp{B378}{1996}{255}.} which allows a comparison to
recent determinations of parton distributions~\ABFR.~\foot{The AB
scheme is a variant of the \MS-HVBM scheme in polarized DIS which
adopts a different subtraction in the definition of the polarized
gluon coefficient function. The scheme is constructed in terms of the
\MS\ scheme in such a way
that only the $z_{qg}$ entry of the scheme change matrix is
nonzero. This means that only the $\Delta C_g^d$ coefficient function
is affected by the scheme change, while all other coefficient
fucntions are the same as in the \MS\ scheme.}
The full Higgs production unpolarized coefficient functions are given
in ref.~\rvirt, while the calculation of their 
polarized counterparts was presented in
the previous section. For completeness, we list in Appendix A
the full set of
NLO polarized and unpolarized coefficient functions in moment 
space in the \MS-AB scheme.

Combined bounds on $\Delta \Sigma(N,Q^2)$ and $\Delta g(N,Q^2)$ can
now be derived using eqs.~\nloqbound-\nlogbound\ for various
scales and values of the moment variable. We will assume everywhere
the equality of the
factorization  and renormalization scales; specifically
in eq.~\nlresxg,\nlresxq\
$\mu^2=m_h^2$. 
The bounds are only
significant for $N>1$, because the first moments of unpolarized
distributions diverge. Furthermore, the scale has to be chosen large
enough that the NL truncation of perturbation theory is a reasonable
approximation. On the other hand, the bounds will only be relevant at
not too large values of the scale, since at asymptotically large scales all
bounds are trivially satisfied because of the asymptotic vanishing,
eq.~\limev, of
the ratios $\Delta \Sigma\over
\Sigma$ and $\Delta g\over g$. Furthermore, at
asymptotically large $Q^2$, NLO corrections are negligible and all
bounds reduce to the naive LO ones discussed in Sect.~2. 

\topinsert
\vbox{\hfil\hbox{\epsfxsize=6.truecm\epsfbox{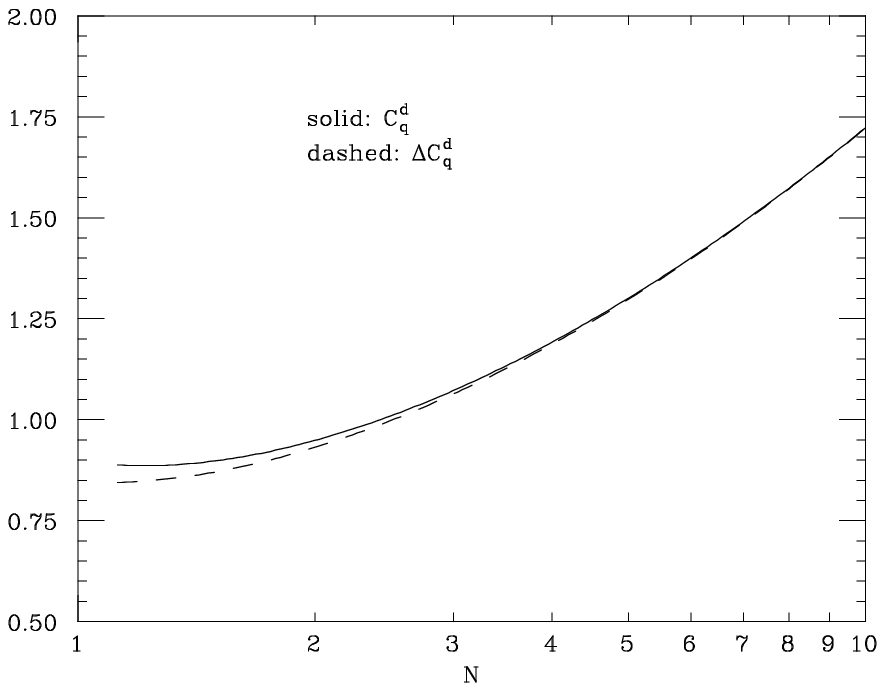}\hskip1truecm
\epsfxsize=6.truecm\epsfbox{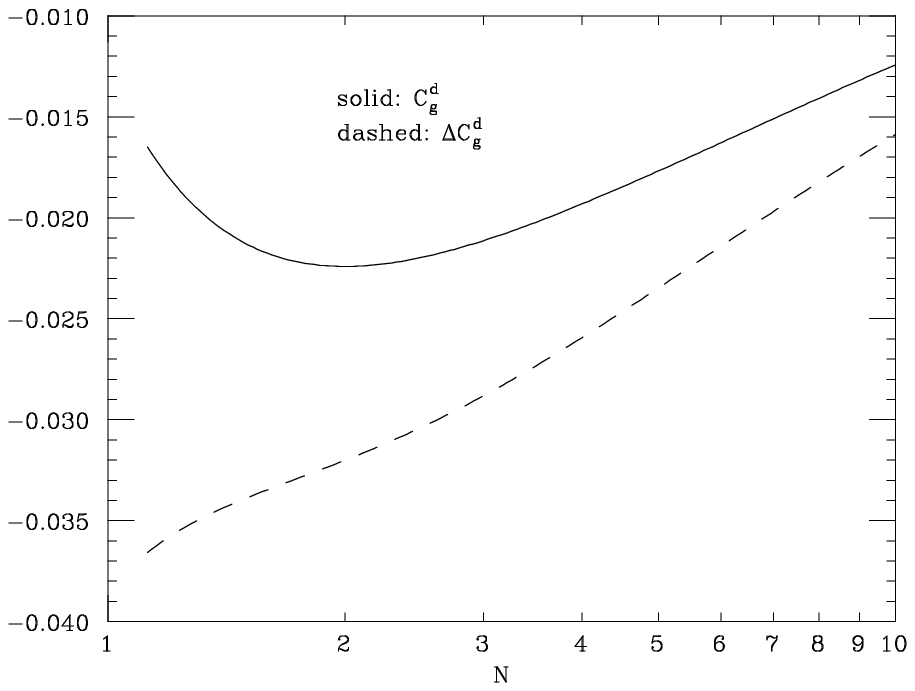}}\smallskip
\hfil\hbox{\epsfxsize=6.truecm\epsfbox{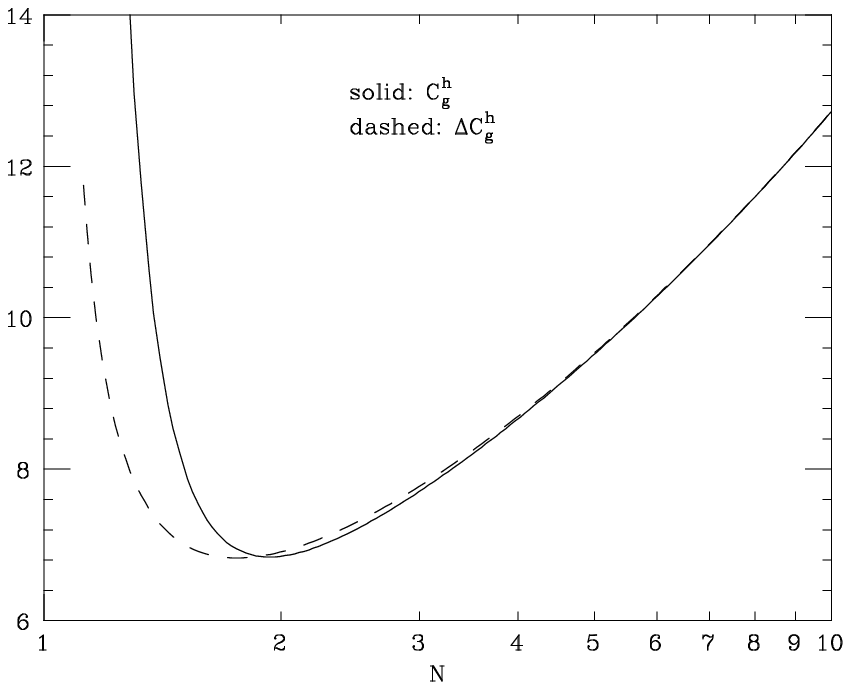}\hskip1.truecm
\epsfxsize=6.truecm\epsfbox{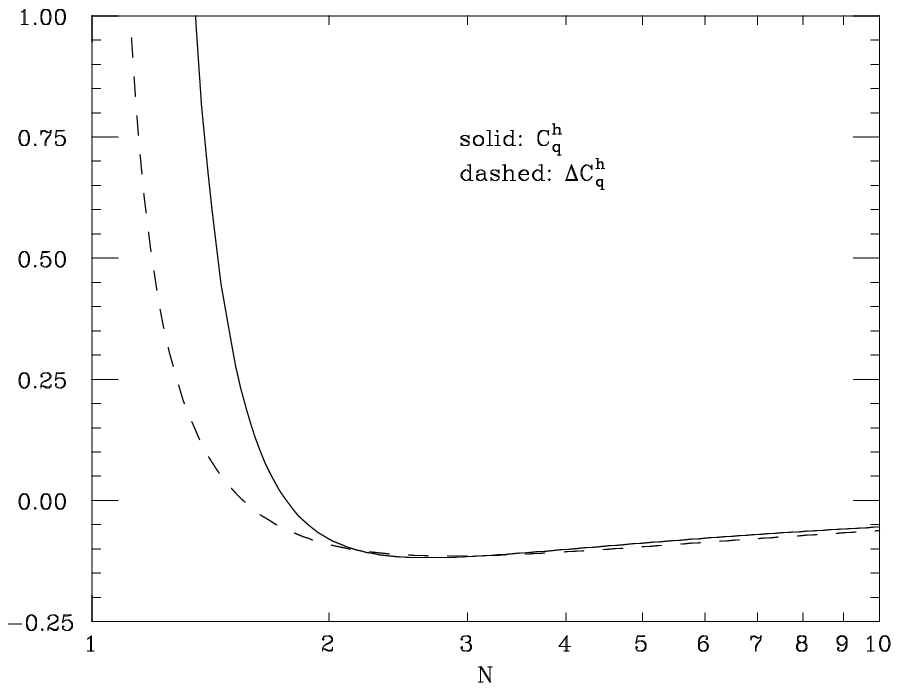}}\hfil}
\bigskip\noindent{\footnotefont\baselineskip6pt\narrower
Figure 2: The eight NLO coefficient functions, evaluated at
$Q^2$=1~GeV$^2$, as  functions of the
moment variable $N$.
\medskip}
\endinsert
In fig.~2 the eight NLO  coefficient functions
  are plotted for $Q^2=1$~GeV$^2$
as functions of the moment variable. 
It is interesting to note that 
the NLO corrections to the Higgs production process in the gluon
channel (i.e. to $\Delta C^h_g$ and $C^h_g$)
are very large (as large as 100\%  at the Higgs scale, and much larger
at the low scale considered here). 
As pointed out in ref.~\rvirt, this is mostly due to
the very large size of virtual corrections. These corrections,
however, are proportional to the Born process and thus only contribute
to  the gluon channel spin-nonflip cross section, and
indeed, $|\Delta C^h_g- C^h_g|\ll|\Delta C^h_g|+ |C^h_g|$.
Since the NLO corrections to the positivity bounds are only due to the
nonvanishing of the spin-flip process, it follows that the NLO corrections to
the bounds are actually very small and can be safely treated within
perturbation theory despite the large size of the NLO corrections to
the coefficient functions themselves. 

The pattern of deviations from the LO behavior displayed by the
NLO corrections to the
coefficient functions of fig.~2 is rather complicated:
all the Higgs production coefficient functions
satisfy $|\Delta C^{h}|< C^{h}$ for small $N$ but 
$|\Delta C^{h}|>
C^{h}$  when $N$ is large; the DIS coefficient functions for all
$N$ have $|\Delta C^{d}_s|< C^{d}_s$ but 
$|\Delta C^{d}_g|> C^{d}_g$ (notice that both $\Delta C_g$ and $C_g$
are negative). The fact
that some NLO corrections to the
polarized coefficient functions are larger than their
unpolarized counterparts means that some of the NLO 
contributions to the associate cross
sections are negative.

The bounds eqs.~\nloqbound-\nlogbound\ are useful to
constrain  more  poorly known parton distributions  
by the better
known ones.
Specifically, the bounds are significant for high
values of the moment variable $N$ because this singles out the large
$x$ region, where the unpolarized distribution is reasonably
constrained by the data whereas very little is known about the
polarized distribution.

\topinsert
\vbox{\hfil\epsfxsize=9.truecm\epsfbox{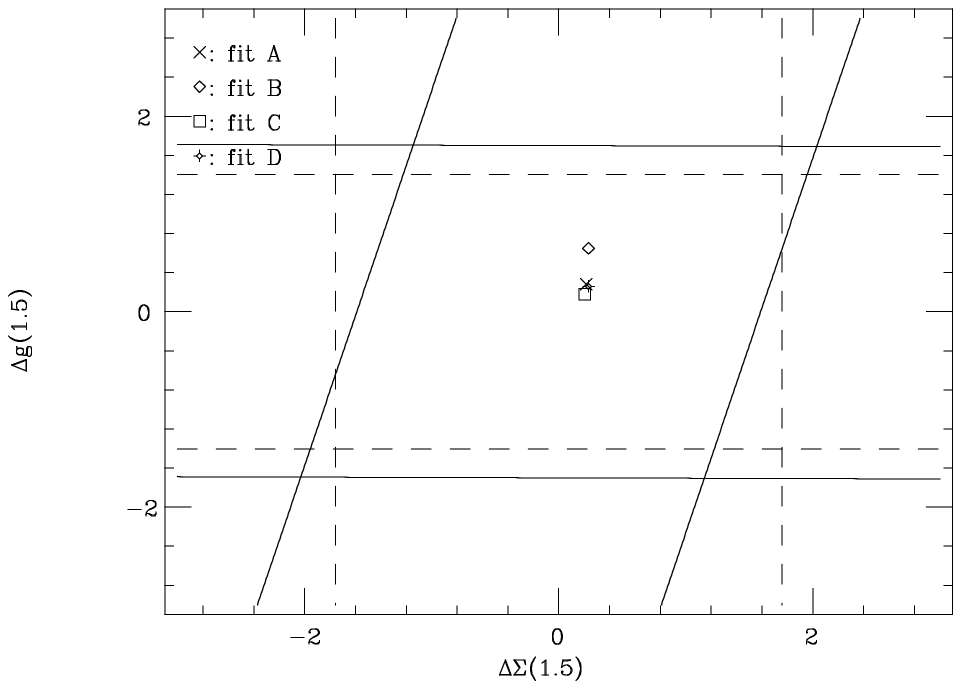}\hfil\medskip
\hfil\epsfxsize=9.truecm\epsfbox{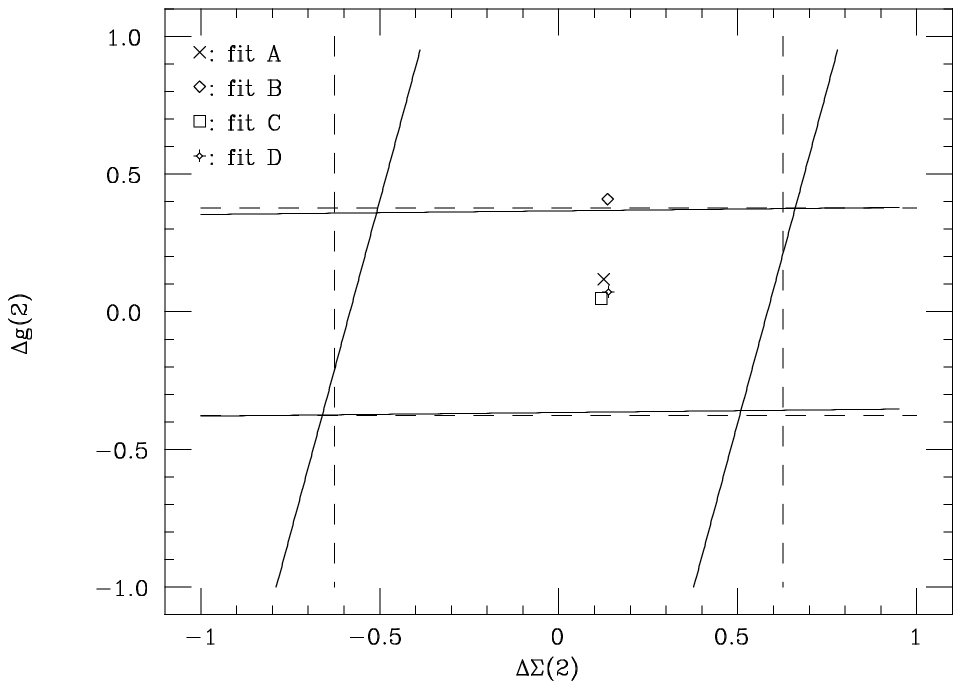}\hfil\medskip
\hfil\epsfxsize=9.truecm\epsfbox{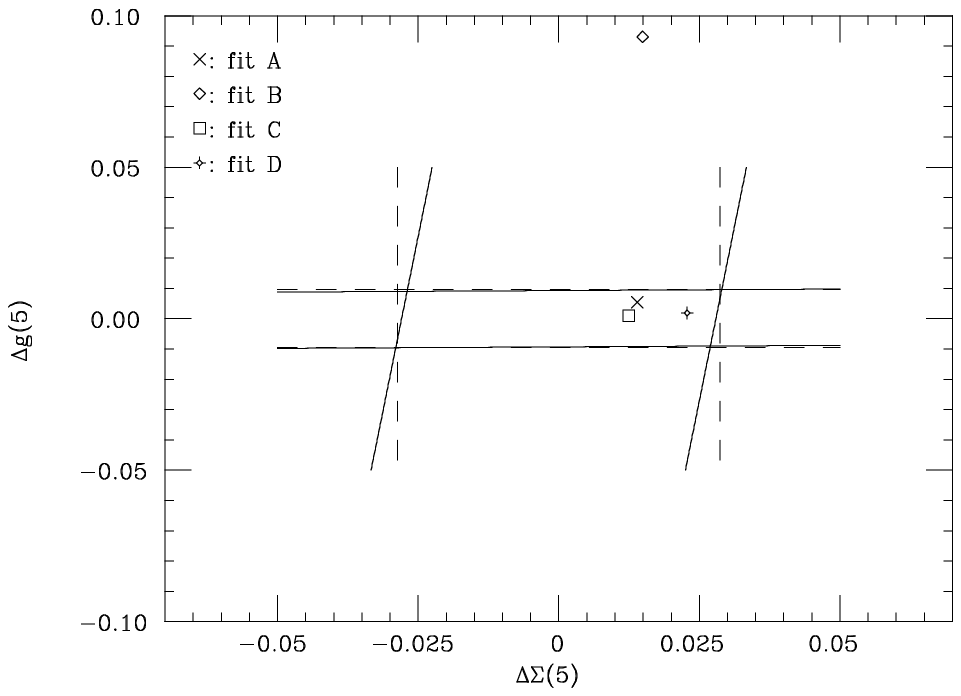}\hfil\medskip
\bigskip\noindent{\footnotefont\baselineskip6pt\narrower
Figure 3: The LO (dashed lines) and NLO (solid lines) positivity
bounds on $\Delta \Sigma(N)$ and $\Delta g(N)$ for $Q^2=1$~GeV$^2$ and
$N=1.5,\,2,\,5$.
The values of $\Delta \Sigma$ and $\Delta g$ corresponding to the  NLO
fits  to $g_1$ data
of ref.~\ABFR\ are also shown.
\medskip}}
\endinsert
In fig.~3 we present the LO and NLO  bounds
eqs.~\nloqbound-\nlogbound\ 
for several relevant values of
the moment variable $N$. The bounds are given 
in the form of an allowed region in the $\Delta \Sigma$--$\Delta g$
plane. The contours depend on the unpolarized
distributions, which are taken from the CTEQ4LQ set~\ref\cteq{H.~L.~Lai et
al., {\it Phys. Rev.}
{\bf D55} (1997) 1280.}.
The maximal allowed  values of the unpolarized
distributions are thus affected by the uncertainity on the unpolarized
distributions.
The size of the uncertainty of parton
distributions
cannot be evaluated precisely at present.  However, in the central $x$
region ($0.01\lsim x\lsim 0.8$) the uncertainty
can be considered to be essentially negligible on the unpolarized
quark distribution, while a recent analysis~\ref\cteqg{H.~L.~Lai et
al., {\tt hep-ph/9801444}.} suggests that the
uncertainty   on the unpolarized gluon distribution is of order $10
\%$
unless $x$ is very small ($x\lsim 10^{-3}$) or large ($x\gsim 0.2$) the
gluon being vanishingly small at large $x$. This should give an idea
on the expected uncertainty on the bounds of fig.~3 due to the errors
on parton distributions, while the
difference between LO and NLO bounds gives an idea of the uncertainty
related to higher order corrections.

 The bounds are presented at  the scale
$Q^2=1$~GeV$^2$ which is often chosen as ``initial'' scale in
parametrization of parton distributions, and which can be taken as a
lower limit for the applicability of perturbation theory. The NLO
corrections are
seen to modify and distort the allowed area in the
$\Delta\Sigma$-$\Delta g$  plane. The pattern of modification of the
LO bounds follows the pattern of NLO corrections to the coefficient
functions displayed in fig.~2. In particular, it is clear that
if $|\Delta C^{(1)}(N)|\leq C^{(1)}(N)$ (i.e. the NLO contributions to the
various cross sections are all positive)
 then the NLO bound is
less restrictive than the LO bound. For instance the quark bound becomes
\eqn\lrbounda{\eqalign{&-{\left[C^d_s(N,Q^2)  \Sigma(N,Q^2) + 
2n_f\left(C^d_g(N,Q^2)   g(N,Q^2)+\Delta C^d_g(N,Q^2)   
\Delta g(N,Q^2)\right)\right]\over
\Delta 
C^d_s(N,Q^2)}
\cr&\qquad \qquad\qquad \qquad\qquad \qquad
\leq\Delta \Sigma(N,Q^2)\leq \cr&
{\left[C^d_s(N,Q^2)  \Sigma(N,Q^2) + 
2n_f\left(C^d_g(N,Q^2)   g(N,Q^2)-\Delta C^d_g(N,Q^2)   
\Delta g(N,Q^2)\right)\right]\over
\Delta C^d_s(N,Q^2)}.\cr}}
This is always less restrictive than
\eqn\lrboundb{|\Delta \Sigma(N,Q^2)|\leq 
{C^d_s(N,Q^2)  \over
\Delta C^d_s(N,Q^2)}\Sigma(N,Q^2) }
because
$C^d_g(N,Q^2)   g(N,Q^2)\pm\Delta C^d_g(N,Q^2)   
\Delta g(N,Q^2)$ is always positive if $|\Delta C^d_g|\leq C_g^d$
since to NLO accuracy we can use
eq.~\gbound\ for the gluon contribution. Eq.~\lrboundb\ is
manifestly less restrictive than the LO bound eq.~\sibound\ if 
$|\Delta C^d_s|\leq C_s^d$. If instead $|\Delta C^d_s|> C_s^d$ the NLO
bound will be more restrictive if the gluon contribution can be
neglected (so that eq.~\lrboundb\ holds) but could still be more
restrictive than the LO one if the gluon contribution is large.
Similar arguments apply to the gluon
distribution.

Inspection of fig.~3 shows that at $Q^2=1$~GeV$^2$
 the NLO corrections to the LO bounds turn out to be
reasonably small, but non-negligible
for small $N$--moments, where they tend to make the bound less
 restrictive than the LO bound if both $\Delta\Sigma$ and $\Delta g$
are positive (as experiments appear to indicate).
The corrections are instead essentially negligible for large values of
 $N$.
However, at sufficiently low
scale all the NLO corrections will start growing and dramatically modify
the LO bounds, but use of perturbation theory at such scales is
not advisable.

Figure~3 also shows the values of $\Delta \Sigma$ and
$\Delta g$ obtained in a recent determination~\ABFR\ of polarized
parton distributions. Different fits correspond to different choices
of functional forms and theoretical assumptions on the parton
distributions, and give a feeling for the spread of the results.
It is apparent that one fits (fit B) display large percentage violations of
the positivity bound on $\Delta g(N)$  
for large values of the moment variable $N$.
This reflects the fact that large values of 
$N$ correspond to the large $x$ region where $\Delta g(N)$
is extremely small and the fitted results are
 affected by very large
uncertainties. Indeed, the main purpose of ref.~\ABFR\ was instead
to
provide a reliable determination of the first moments of $\Delta
\Sigma$ and $\Delta g$, while the large $x$ i.e. large $N$ behavior
of the parton distributions could not be determined
accurately. This however shows explicitly how positivity bounds can be
used to improve a determination of parton distributions, and
specifically $\Delta g$ by providing
independent information. Indeed, we could modify fit B in such a way
that the bounds are respected, without significantly
affecting the quality of the
fit to data and the determination of the first moment. However,
imposing that the bound is satisfied we would 
get a fit with reduced
uncertainity on $\Delta g$ in the large $x$ region.

\newsec{Summary and outlook}
In the naive parton model the polarized parton densities in a proton
with
helicity +1/2, $p_+$ and $p_-$, are interpreted as probability
densities
and, as a consequence, are positive semidefinite. Thus, given $\Delta
p=p_+-p_-$ and $p=p_++p_-$, they satisfy $|\Delta p|\leq p$, for any
$p=q_i,
\bar q_i, g$. The QCD-improved parton model reproduces the naive
parton
results at LO, i.e. when all higher order perturbative corrections (as
well
as all power corrections) are neglected. In this limit, if the naive
positivity bounds  for parton densities were violated, one could
exhibit hard processes with unacceptable negative cross-sections. For
quarks obvious examples of such processes are given by the polarized
asymmetries $A_1=g_1/F_1$ eq.~\aone. For gluons, the selection of the
process which leads to the bound $|\Delta g| \leq g$ is less direct. One
needs a process whose rate on a polarized proton with helicity +1/2
starts
with $g_+$ and does not contain either $g_-$ or $q_{\pm}$ or $\bar
q_{\pm}$. The process $g+p\rightarrow H+X$, where $H$ is a neutral,
colourless massive scalar (for example, a Higgs boson) satisfies the
required properties.

As is well known, most naive parton model properties and sum rules are in
general violated if higher order QCD corrections are included. For
example, the charge sum rules $u(1,Q^2)-\bar u(1,Q^2)=2$ and similar
ones
[where $u(N,Q^2)$ is the N-th moment of the u-quark density $u(x,Q^2)$]
are
in general violated, unless the  unpolarized quark densities are
defined by the structure function $F_2$,\ref\partsch{G.~Altarelli,
R.~K.~Ellis and 
G.~Martinelli, {\it  Nucl. Phys.} {\bf B143} (1978) 521; {\bf B157}
(1979) 461.}  i.e. by extending to all
orders,
by definition, the LO expression of $F_2$ in terms of quark
densities.
Similarly also the momentum sum rule or the LO positivity conditions
are
in general violated at NLO and higher accuracy, with corrections which
depend on the definitions of the relevant quantities beyond LO.

In this article we have computed the NLO corrections to the
positivity relations
in
a set of specified definitions. In particular we adopted
the \MS\ scheme and a simple set of physical processes for the
definition of parton densities, including the above mentioned Higgs
production process for the gluon definition. Of particular interest in
practice are the positivity bounds in the polarized singlet sector
$\Delta
\Sigma$, $\Delta g$. This is because the determination of the
polarized gluon
density attracts considerable attention at present. When
fitting the
data
on scaling violations often the gluon comes close to the LO positivity
bound at
large $x$, so that the precise form of the bound at NLO can be
relevant. While at LO
the positivity bounds in the singlet sector are given separately for
$\Delta
\Sigma(N)$ and $\Delta g(N)$ ($|\Delta \Sigma(N)|\leq\Sigma(N)$,
$|\Delta
g(N)|\leq g(N)$), at NLO due to quark-gluon mixing the allowed domain
is
modified
into a more complicated region of the $\Delta \Sigma - \Delta g$
plane. 

Our
detailed quantitative results show that the modified region 
in the $\Delta \Sigma - \Delta g$ plane can be either more or
less restrictive than the LO positivity domain, depending on the
values of N,
$\Sigma(N)$ and $g(N)$. Interestingly, we also find that the
corrections to the
positivity limiting contours are in general much smaller than the size
of the
corrections on the individual coefficient functions. In particular
this is true
for the corrections to the Higgs process. It is well known that the
NLO
corrections to Higgs production by gluon-gluon fusion are quite
large~\rvirt. But the
bulk of the corrective terms is from virtual diagrams which, being
proportional
to the Born term, do not modify the LO bounds. 

As a consequence, we
find that,
even at relatively small values of $Q^2$, such as $Q^2\sim1$~GeV$^2$, the
modified
domains are reasonably close to the corresponding LO region. Of course
if $Q^2$
is further lowered down to a clearly non--perturbative region, the NLO
corrections
eventually explode and the NLO positivity contours become very
different than
the
LO contours. It is important to note that if the LO positivity bounds
are
imposed
at very small $Q^2$ they in general produce exaggerately constraining
limits on
the parton densities. This is because the evolution goes in the
direction of
making the positivity bounds trivial at $Q^2\rightarrow\infty$.
Hence, applying the LO bounds at very small  $Q^2$ leads to an evolved
domain at larger $Q^2$ that is considerably smaller than the LO bound
at the larger value of $Q^2$. It follows that if parton distributions are
parametrized at a low scale, positivity bounds can only be safely imposed
at a higher scale, where perturbation theory is reliable. 
A possible strategy is to use positivity bounds to discard fitted
parton distributions which do not satisfy them, thereby reducing the 
uncertainty on the results of the fit.

Imposing positivity bounds consistently at NLO in a given scheme will
then guarantee positivity of physical cross sections for the defining
processes. It will also guarantee
positivity
of any other process whose LO coefficient functions do not
have the partonic form eqs.~\locfq,\locfg, because the LO bounds for
such processes will be less restrictive. It will not, however,
guarantee positivity of other processes for which eqs.~\locfq,\locfg\
hold, because then positivity would depend on the relative sign and
size of NLO corrections. 

In conclusion, positivity bounds on parton distributions
derived from specific defining processes
are in general useful as necessary
conditions which provide complementary information on polarized parton
distributions, especially at large $x$.
\medskip
{\bf Acknowledgements:} We thank M.~Anselmino,
R.~D.~Ball and M.~Spira for
discussions and correspondence. This work was supported in part by the
EU Fourth Framework Programme
``Training
and Mobility of Researchers'', Network `Quantum Chromodynamics and the Deep
Structure of Elementary Particles', contract FMRX-CT98-0194 (DG 12--MIHT).

\listrefs

\vfill
\eject

\appendix{A}{Coefficient functions in moment space}
We list here the Mellin moments $C(N)\equiv \int_0^1 x^{N-1} C(x)\, dx$
of the NLO contributions to the
coefficient functions in the
\MS\ scheme for
generic values of the ratio of factorization and renormalization
scales $k\equiv{\mu^2_R\over\mu^2_F}$. The full coefficient functions
are defined in eq.~\cexp; the LO contributions are given in
eqs.~\locfq,\locfg. 

\medskip\noindent{\it Deep-inelastic scattering NLO coefficient functions}:
\eqnn\nlcdq\eqnn\nlcdg\eqnn\nldcdq\eqnn\nldcdg
\smallskip
$$\eqalignno{C^{d,\,(1)}_S(N)&=
C_F\bigg[S_1(N-1)
      \left(S_1(N+1)+{3\over 2}\right)
      +{4\over N}-{1\over {N+1}}-{9\over 2}\cr&
\qquad\qquad-S_2(N-1)
   \bigg]+P_{qq}\ln k    &\nlcdq\cr
C^{d,\,(1)}_g(N)&=
- T_R \left[S_1(N-1) {{N^2+N+2}\over {N(N+1)(N+2)}}+{1\over {N+2}}\right]
  \cr&\qquad        +P_{qg}\ln k  
                  &\nlcdg\cr
\Delta C^{d,\,(1)}_S(N)&=
C_F \bigg[S_1(N-1) \left({3\over 2}+S_1(N+1)\right)
\cr&\qquad\qquad
-S_2(N-1)-{9\over 2}+{3\over N}\bigg]+\Delta P_{qq}(N) \ln k
                   &\nldcdq\cr
\Delta C^{d,\,(1)}_g(N)&=-T_R  {{N-1}\over {N(N+1)}} 
\left[S_1(N-1)+1\right]
+\Delta P_{qg}(N) \ln k . &\nldcdg\cr}$$
The AB-\MS\ scheme~\bfr\ (used in sect.~5) is obtained adding
to $\Delta C^{d,\,(1)}_S(N)$ eq.~\nldcdg\
a contribution equal to $-T_R\over N$, while all
other coefficient functions are left unchanged.
\medskip
\noindent{\it Higgs production NLO coefficient functions}:
\smallskip
\eqnn\nlchq\eqnn\nlchg\eqnn\nldchg\eqnn\nldchq
$$\eqalignno{C^{h,\,(1)}_g(N)&=C_A\left[{11+4\pi^2\over3}
  -{22\over {N(N^2-1)(N+2)}}\right]\cr&\qquad\qquad\qquad+2Q_{gg}(N) 
          +2 P_{gg}(N)\ln k
      &\nlchq\cr
C^{h,\,(1)}_S(N)&=C_F{N^2-N-3\over N(N^2-1)}
+Q_{gq}(N)+ 
P_{gq}(N)\ln k  &\nlchg\cr
\Delta C^{h,\,(1)}_g(N)&=C_A\left[{11+4\pi^2\over3}
  +{22\over {N(N^2-1)(N+2)}}\right]\cr&\qquad\qquad\qquad+2\Delta Q_{gg}(N)
          +2 \Delta P_{gg}(N)\ln k&\nldchg\cr
\Delta C^{h,\,(1)}_S(N)&=C_F{3\over N(N^2-1)}+ \Delta Q_{gq}(N)
     +\Delta P_{gq}(N)\ln k .             &\nldchq\cr}$$
\smallskip
\noindent
We have 
defined\eqnn\sdef\eqnn\qgdef\eqnn\dqgdef\eqnn\qqdef\eqnn\dqqdef
$$\eqalignno{&S_i(N)=\sum_{k=1}^{N} {1\over k^i}&\sdef\cr
&Q_{gg}(N)=
2 C_A\left[S_1^2(N+2)-2S_1^2(N+1)\right.\cr&\qquad\qquad\qquad\left.
+3S_1^2(N)-2S_1^2(N-1)+S_1^2(N-2)\right]
&\qgdef\cr
\Delta
&Q_{gg}(N)=2C_A\left[2S_1^2(N+1)-3S_1^2(N)+2S_1^2(N-1)\right]&\dqgdef\cr
&Q_{gq}(N)= 
C_F \left[-{{4S_1(N-1)}\over N-1}
                        +{{4S_1(N)}\over{N}}
                        -{{2S_1(N+1)}\over{N+1}}
\right.\cr&\qquad\qquad\qquad\qquad\left.+
{2\over{(N-1)^2}}-{2\over{N^2}}+{1\over{(N+1)^2}}
\right]
 &\qqdef\cr
&\Delta Q_{gq}(N)=C_F\left[-{{4 S_1(N)}\over {N}}
      +{2{S_1(N+1)}\over {N+1}}
      +{2\over {N^2}}-{1\over {(N+1)^2}}\right]
&\dqqdef\cr}$$
Furthermore, $P_{ij}(N)$ and  $\Delta P_{ij}(N)$
are the usual LO polarized and unpolarized anomalous dimensions, 
i.e. the Mellin
transforms of the LO QCD splitting functions, whose
expressions we list here for completeness:
\eqn\splftn{\eqalign{
P_{qq}&=C_F\left[{3\over 2}-{1\over N}-{1\over {N+1}}-2 S_1(N-1)\right]\cr
P_{qg}&= T_R {{N^2+N+2}\over{N(N+1)(N+2)}}\cr
P_{gq}&=C_F{N^2+N+2\over N (N^2-1)}\cr
P_{gg}&=2C_A\left[{1\over N(N-1)}+{1\over (N+1)(+2)}-S_1(N)\right]+
{\beta_0\over2}\cr
\Delta P_{qq}&=C_F\left[{3\over 2}-{1\over N}-{1\over {N+1}}-2
S_1(N-1)\right]\cr 
\Delta P_{qg}&= T_R {{N-1}\over{N(N+1)}}\cr
\Delta P_{gq}&=C_F {N+2\over N(N+1)}\cr
\Delta P_{gg}&=2C_A\left[{2\over N(N+1)}-S_1(N)\right]+{\beta_0\over 2}.\cr
}}

\vfill\eject\bye